\pdfoutput=1
\documentclass[11pt]{article}
\usepackage{listings}
\usepackage{acl}

\usepackage{times}
\usepackage{latexsym}

\usepackage[T1]{fontenc}
\usepackage[utf8]{inputenc}

\usepackage{microtype}

\usepackage{inconsolata}

\usepackage{graphicx}
\usepackage{subcaption}
\usepackage{hyperref}
\usepackage{amsmath}
\usepackage{cleveref}
\usepackage{amsfonts}
\usepackage{enumitem}
\usepackage{booktabs}
\title{KLIPA: A Knowledge Graph and LLM-Driven \\ QA Framework for IP Analysis}

\author{Guanzhi Deng$^1$, Yi Xie$^2$, Yu-Keung Ng$^1$, Mingyang Liu$^1$, Peijun Zheng$^1$, \\\textbf{Jie Liu$^3$, Dapeng Wu$^1$, Yinqiao Li$^1$, Linqi Song$^1$} \\
  $^1$City University of Hong Kong \quad $^2$Integrated Global Solutions Limited \\\ $^3$North China University of Technology \\
  \texttt{guanzdeng2-c@my.cityu.edu.hk, linqi.song@cityu.edu.hk}
}

\begin{document}
\maketitle
\begin{abstract}
Effectively managing intellectual property is a significant challenge. Traditional methods for patent analysis depend on labor-intensive manual searches and rigid keyword matching. These approaches are often inefficient and struggle to reveal the complex relationships hidden within large patent datasets, hindering strategic decision-making. To overcome these limitations, we introduce KLIPA, a novel framework that leverages a knowledge graph and a large language model (LLM) to significantly advance patent analysis. Our approach integrates three key components: a structured knowledge graph to map explicit relationships between patents, a retrieval-augmented generation (RAG) system to uncover contextual connections, and an intelligent agent that dynamically determines the optimal strategy for resolving user queries. We validated KLIPA on a comprehensive, real-world patent database, where it demonstrated substantial improvements in knowledge extraction, discovery of novel connections, and overall operational efficiency. This combination of technologies enhances retrieval accuracy, reduces reliance on domain experts, and provides a scalable, automated solution for any organization managing intellectual property, including technology corporations and legal firms, allowing them to better navigate the complexities of strategic innovation and competitive intelligence.

% Experimental results show that it yields an average of \textcolor{red}{XXX} on top of a \textcolor{red}{XXX} implementation. Also, \textcolor{red}{XXX} (interesting findings). 
\end{abstract}

\section{Introduction}

Managing patents effectively is a critical challenge for a diverse array of entities, including technology corporations, specialized law firms, universities, and research institutions. Conventional approaches rely heavily on manual keyword searches across multiple patent databases (e.g., Google Patents, USPTO, etc.), followed by labor-intensive filtering and rigid categorization, making it difficult to identify meaningful connections between patents. These inefficiencies are further exacerbated by the growing volume of patent data. According to statistics from the European Patent Office (EPO), comprehensive patent searches involve 1.3 billion technical records across 179 databases, with approximately 600 million documents processed monthly, leading to an increasing demand for well-trained manpower and extended processing times\footnote{\url{https://link.epo.org/web/EPO_Strategic_Plan_2023_en.pdf}}.

\begin{figure*}[ht]
  \centering
  \includegraphics[width=1.00\textwidth]{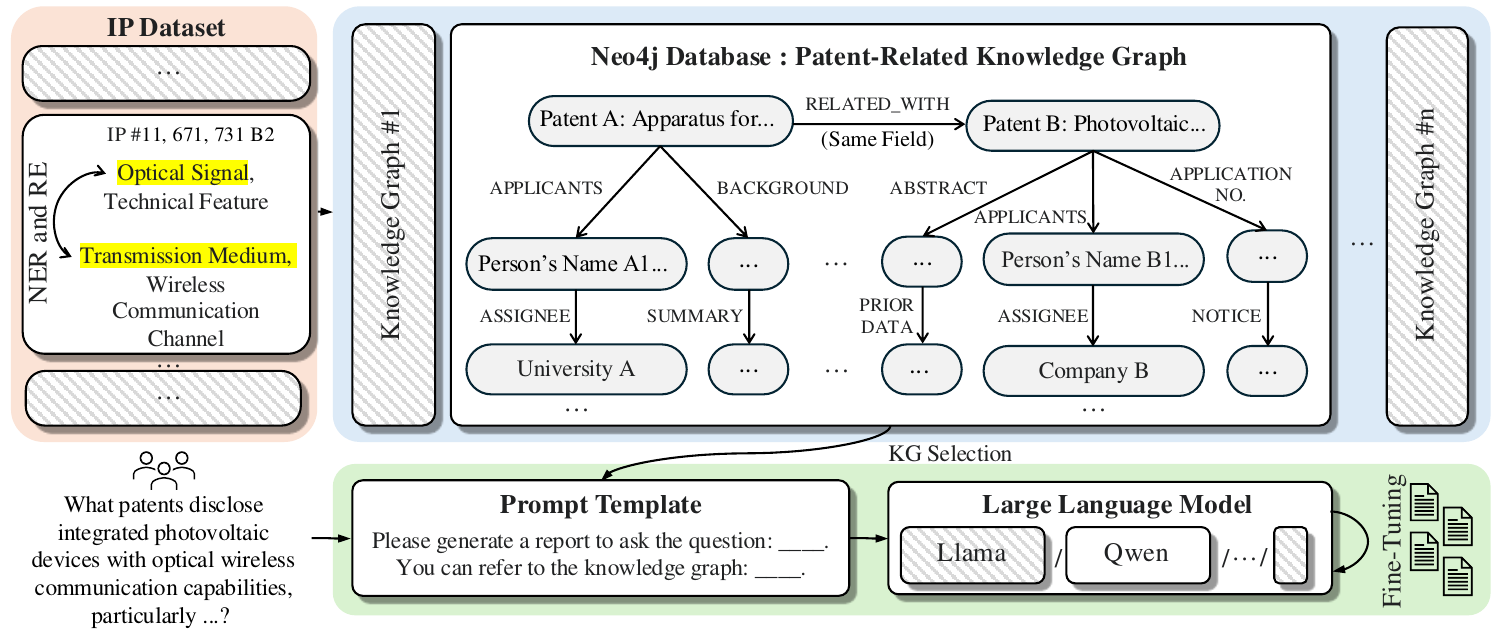}
  \caption{Two-phase workflow of KLIPA: Phase 1 (Knowledge Graph Construction) parses multilingual patent documents, extracts technical entities and their relationships, and builds a structured network in a graph database. Phase 2 (QA Agent Deployment) implements a hybrid retrieval mechanism on the graph, combining semantic matching (via RAG) with relational reasoning (via KG). The system adopts a unidirectional data flow: the knowledge graph serves as the foundational layer, while the Agent generates targeted responses through reasoning steps tailored for user queries. Modules are decoupled via standardized interfaces, allowing independent updates to both the graph and QA models.}
  \label{fig:framework}
\end{figure*}

% The traditional patent retrieval process is inefficient, error-prone, and highly dependent on domain knowledge. According to statistics from the European Patent Office (EPO), comprehensive patent searches utilize 1.3 billion technical records across 179 databases, and approximately 600 million documents processed in monthly patent searches further increase the demand for well-trained manpower and time\footnote{\url{https://link.epo.org/web/EPO_Strategic_Plan_2023_en.pdf}}. 

To address these challenges, AI-driven patent retrieval methods have been explored\cite{shomee2024comprehensivesurveyaibasedmethods}. For example, the Chemical Abstracts Service (CAS) and the National Institute of Industrial Property of Brazil (INPI) have collaborated on AI-based search workflows which reduce processing time by 77\%\footnote{\url{https://www.cas.org/resources/cas-insights/ai-proves-effective-improving-patent-office-efficiency}}. Additionally, \citet{AlthammerHH21} applied BERT-PLI as a foundation to enhance patent retrieval. However, existing solutions suffer from several key limitations: 

\begin{itemize}[nosep,leftmargin=*]
\item[$\bullet$] Many solutions are proprietary commercial tools, offering limited transparency and accessibility. 
\item[$\bullet$] Open-source neural models remain dependent on human experts to filter and reorganize results manually, limiting automation efficiency.
\end{itemize}

% systems are either closed-source commercial solutions without accessible technical details and data, or weak neural models that still require experts to manually filter and reorganize search results. 

To overcome these limitations, we propose the \textbf{K}nowledge Graph and \textbf{L}LM-Driven Question-Answering Framework for \textbf{IP} \textbf{A}nalysis, or KLIPA (\Cref{fig:framework}), to enhance relationship identification, patent retrieval, and knowledge discovery. KLIPA integrates three key components to optimize the retrieval and analysis process: 

\begin{itemize}[nosep,leftmargin=*]
\item[$\bullet$] Patent Knowledge Graph (KG): Captures structured relationships between patents to improve retrieval efficiency. 
\item[$\bullet$] Retrieval-Augmented Generation (RAG): Retrieves semantically relevant patent information and uncovers hidden connections beyond explicit entity relationships.
\item[$\bullet$] ReAct Agent Framework: Dynamically generates reasoning steps during query resolution, determining whether to use KG, RAG, or both for optimal response generation. 
\end{itemize}

% We call our method \textbf{K}G and \textbf{L}LM-Driven QA framework for \textbf{IP} \textbf{A}nalysis (or KLIPA for short). Here, we rely on KG to identify multi-dimensional patent connections through its extensive patent coverage, rely on cutting-edge LLMs to understand queries, and rely on retrieval-augmented generation (RAG) techniques to generate analysis reports. Additionally, a novel hybrid search mechanism is proposed to further explore the inter-connections within patent KGs, with the potential application in other RAG systems. The contributions of this work are three-fold. 

By intelligently combining structured and unstructured retrieval mechanisms, KLIPA significantly enhances retrieval accuracy, reduces reliance on domain experts, and improves patent knowledge discovery. Our main contributions are three-fold:

\begin{itemize}[nosep,leftmargin=*]
\item[$\bullet$] \textbf{A hybrid patent retrieval framework integrating KGs, LLMs, and RAG, enabling both explicit and implicit relationship discovery.} 
\item[$\bullet$] A ReAct-based query resolution mechanism that dynamically determines the best retrieval strategy for each user query, improving adaptability and precision in patent analysis.
\item[$\bullet$] A validated, scalable system for patent management, tested on a university patent database, demonstrating improved knowledge extraction, search accuracy, and reduced manual effort. 
\end{itemize}

Our code is available at the anonymous link\footnote{\url{https://github.com/gz-d/patent_kg}}. We plan to release our constructed patent KG (with more than 1000 patents) in the camera-ready version of this paper, to avoid any disclosure of identifiable information.

\section{Related Work}

% \subsection{Knowledge Graph}
Knowledge graphs are vital for knowledge representation. Early methods relied heavily on rules and feature engineering and therefore faced scalability issues. The Semantic Web by DBpedia \cite{auer2007dbpedia}, YAGO \cite{10.1145/1242572.1242667}, and Freebase \cite{bollacker2007freebase}, allowed knowledge graphs to scale up. Refinement methods combining rules and machine learning improved data quality \cite{paulheim2016knowledge}. Embedding techniques enhanced semantic reasoning \cite{nickel2016holographic}, while GCNs and Transformers boosted semantic search \cite{zhang2019long}. End-to-end frameworks enabled dynamic updates \cite{yao2019kg}.

Term hierarchies improved patent text similarity \cite{li2020computing}, and syntactic features enhanced engineering knowledge graphs \cite{siddharth2022engineering}. Unsupervised language models improved recall but lacked deep understanding \cite{zuo2021patent}. Recent advances with pre-trained models enhanced semantic modeling, with Sentence-BERT and TransE \cite{siddharth2022enhancing} as an example. Automated construction and reasoning benefited from GPT models \cite{trajanoska2023enhancing, heyi2023research, DBLP:journals/corr/abs-2111-11295}, with improved long-text understanding using GPT-4 and Pat-BERT \cite{caikeresearch}. KnowGPT proposed enhanced key knowledge extraction via context-aware prompting \cite{zhang2024knowgpt}.

% \subsection{LLM-based Agent of Knowledge Graph}
Recent works have explored integrating LLMs with KGs to enhance reasoning capabilities. One approach combines LLMs and KGs for legal article recommendations \cite{chen2024leverageknowledgegraphlarge}, while another improves AI-driven legal assistants' reliability by incorporating expert models with LLMs and KGs \cite{cui2024chatlawmultiagentcollaborativelegal}. Autonomous agent frameworks like KG-Agent use LLMs and KGs for more efficient reasoning with reduced resource usage \cite{jiang2024kgagentefficientautonomousagent}. Additionally, R2-KG \cite{jo2025r2kggeneralpurposedualagentframework} introduces a dual-agent framework that separates evidence gathering and decision-making, reducing reliance on high-capacity LLMs while enhancing reliability and accuracy. These advancements demonstrate the growing potential of LLM-based agents for complex reasoning tasks with KGs.

\section{Methodology}
To overcome the limitations of traditional patent information retrieval, we introduce \textbf{KLIPA}, a hybrid framework that combines a structured knowledge graph and an LLM agent-based QA system. By integrating structured and unstructured retrieval, KLIPA improves patent search accuracy, enhances knowledge extraction, and reduces reliance on domain experts.

% Detailed theoretical framework is presented in \ref{section: theory}, and the concrete algorithm implementations are provided in \ref{section: pseudocode}.

\subsection{Patent Knowledge Graph Construction}
\label{sec:patent kg}
\begin{figure*}[ht]
  \centering
  \includegraphics[width=0.99\textwidth]{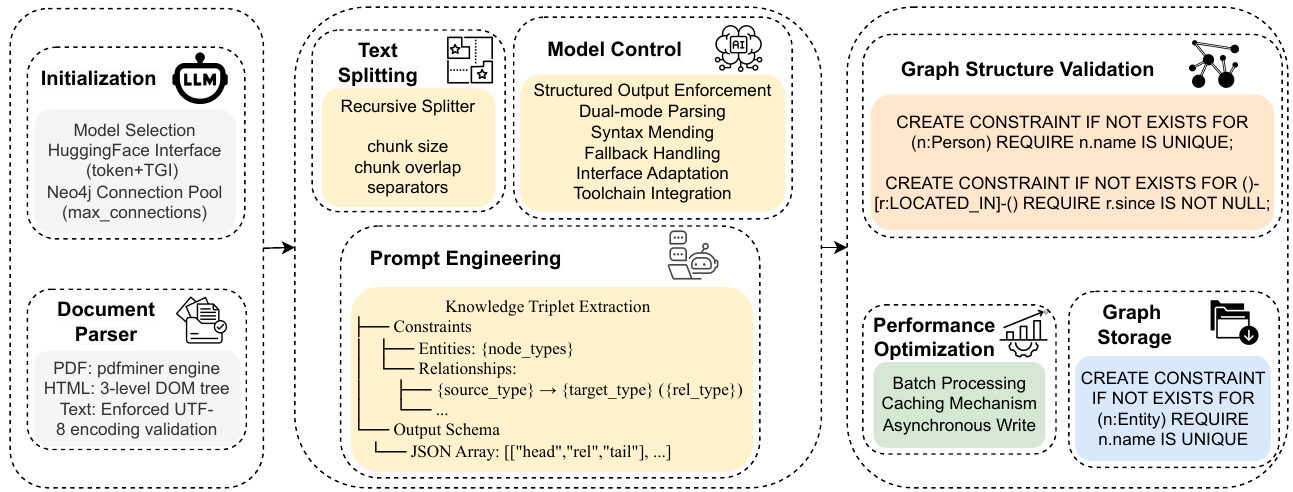}
  \caption{Patent knowledge graph construction pipeline. The process begins with language model initialization and document parsing, followed by text segmentation for efficient knowledge extraction. Structured triplets are extracted using prompt-based methods, ensuring data reliability. Graph validation, caching, and batch processing enhance performance, while the final data is stored in a Neo4j database with uniqueness constraints for consistency.}
  \label{fig:kg}
  \vspace{-6mm}
\end{figure*}

The patent knowledge graph serves as the backbone of KLIPA, transforming unstructured patent documents into structured data that enables efficient retrieval, reasoning, and analytics. As illustrated in \Cref{fig:kg}, the process consists of multiple phases: entity extraction, relationship identification, graph construction, and querying. The following sections formalize these processes using a graph-based architecture for storing and querying patent data.

\paragraph{System Initialization and Data Parsing.}
The pipeline starts with the initialization of core components:
\begin{itemize}[nosep,leftmargin=*]
    \item[$\bullet$] \textit{Model Initialization:} The system selects an appropriate language model and establishes a connection to the Neo4j graph database. Parameters such as token settings and database connection pool limits are configured for optimal performance.
    \item[$\bullet$] \textit{Document Parsing:} Given a set of patent documents \( \mathcal{P} = \{ p_1, p_2, \dots, p_n \} \), the system processes PDFs, HTML, and plain text documents using format-specific parsers. Each document's text content is extracted, ensuring UTF-8 encoding compliance to maintain data integrity.
\end{itemize}

\paragraph{Text Splitting and Preprocessing.}
To facilitate efficient entity and relationship extraction, documents are divided into manageable chunks:
\begin{itemize}[nosep,leftmargin=*]
    \item[$\bullet$] \textit{Recursive Splitting:} Each document \( p_i \) is recursively segmented based on natural language boundaries such as paragraph breaks, punctuation, and line separators. Each resulting text chunk maintains semantic coherence.
    \item[$\bullet$] \textit{Chunk Overlap Mechanism:} To preserve context across segments, overlapping text spans are included between consecutive chunks, enhancing entity continuity and improving relationship extraction accuracy.
\end{itemize}

\paragraph{Formalization of Entity Extraction and Classification.}
Let \( \mathcal{P} = \{ p_1, p_2, \dots, p_n \} \) denote a collection of \( n \) patent documents, where each document \( p_i \in \mathcal{P} \) contains textual content. Denote by \( \mathcal{E}_i \subseteq \mathcal{E} \) the set of entities extracted from the document \( p_i \), where \( \mathcal{E} \) is a predefined set of entity types, such as inventors, technologies, companies, and patent classifications. The extraction function \( E \) maps each document to its corresponding set of entities as follows:
\begin{align}
E(p_i) &= \{ e_{i1}, e_{i2}, \dots, e_{im_i} \} \nonumber, \\
& \forall p_i \in \mathcal{P},\ e_{ij} \in \mathcal{E},\ 1 \leq j \leq m_i,\ m_i \in \mathbb{N}.
\end{align}
where \( m_i \) is the number of entities identified in document \( p_i \). The complete set of entities across all documents is: \(\mathcal{E}_{total} = \bigcup_{i=1}^{n} \mathcal{E}_i \subseteq \mathcal{E}\). Thus, the total number of distinct entity types extracted from the documents is bounded by \( \mathcal{E}_{total} \).

\paragraph{Entity Relationship Extraction.}
Next, we define the extraction of relationships between entities within the patent documents. Let \( \mathcal{R} \subseteq \mathcal{E} \times \mathcal{E} \) represent the set of potential relationships between entity pairs (e.g., invented by, owned by, references). The relationship extraction function \( R \) maps each document and entity pair to a binary value indicating the presence or absence of a relationship:
\begin{equation}
R(p_i, e_i, e_j) = 
\begin{cases} 
1, & \text{if } (e_i, e_j) \in \mathcal{R} \text{ for } p_i, \\
0, & \text{otherwise}.
\end{cases}
\end{equation}
The set of relationships \( \mathcal{R}_i \) extracted from document \( p_i \) is then given by:
\begin{equation}
\mathcal{R}_i = \{(e_i, e_j) \mid R(p_i, e_i, e_j) = 1, \ e_i, e_j \in \mathcal{E}\}.
\end{equation}
The overall relationship graph \( \mathcal{G}_R = (V, E_R) \) is constructed by combining the relationships from all documents, where \( V = \mathcal{E} \) and \( E_R = \bigcup_{i=1}^{n} \mathcal{R}_i \).

\paragraph{Graph Construction in Neo4j Database.}
The knowledge graph constructed from the extracted entities and relationships is stored in a Neo4j-based graph database \( \mathcal{D} \). We define the graph \( \mathcal{G} = (V, E) \), where \( V = \mathcal{E} \) and \( E = \mathcal{R} \). This graph allows for efficient querying and retrieval of information through operations such as neighborhood queries and subgraph extraction. Given an entity \( e_i \in V \), the neighborhood \( \mathcal{N}(e_i) \) is defined as \( \mathcal{N}(e_i) = \{ e_j \mid (e_i, e_j) \in E \} \).

Additionally, for a subset of entities \( S \subseteq V \), the induced subgraph \( \mathcal{G}_S = (S, E_S) \) is defined as:
\begin{equation}
E_S = \{ (e_i, e_j) \mid e_i, e_j \in S, (e_i, e_j) \in E \}.
\end{equation}

\paragraph{Performance Optimization and Data Integrity.}
To improve scalability and maintain data consistency:
\begin{itemize}[nosep,leftmargin=*]
    \item[$\bullet$] \textit{Batch Processing:} Enables bulk data import, reducing overhead in large-scale entity and relationship insertion.
    \item[$\bullet$] \textit{Caching Mechanism:} Enhances retrieval efficiency by storing frequently queried nodes and paths.
    \item[$\bullet$] \textit{Asynchronous Write Operations:} Speeds up data insertion and minimizes system latency.
\end{itemize}

\subsection{LLM-Based QA Agent Implementation}
\label{sec:patent qa}
While the knowledge graph captures explicit entity relationships of patents, it does not account for latent similarities or complex reasoning over multiple data sources. To overcome this limitation, KLIPA integrates an LLM-powered QA agent, which dynamically determines the best retrieval strategy based on user queries. This component fuses:
\begin{itemize}[nosep,leftmargin=*]
    \item[$\bullet$] Agent-driven reasoning to optimize query resolution.
    \item[$\bullet$] Graph-based retrieval for structured searches.
    \item[$\bullet$] RAG for uncovering semantic relationships.
\end{itemize}

The proposed QA agent is composed of several interconnected modules: system initialization and model configuration, document indexing, reasoning and retrieval, and interactive query handling. In the following sections, we formalize these components using a modular architecture and present key mathematical formulations that underpin the system’s functionality.

\paragraph{System Initialization and Model Configuration.}
The agent is instantiated by selecting an appropriate LLM backend and establishing a connection to a Neo4j-based vector database. During initialization, the following steps are executed:
\begin{itemize}[nosep,leftmargin=*]
    \item[$\bullet$] \textit{Model Initialization:} The system instantiates the chosen LLM with model-specific parameters.
    \item[$\bullet$] \textit{Vector Database Connection:} A vector database is configured via the \texttt{Neo4jVector} interface, which employs dense embeddings to index patent documents.
    \item[$\bullet$] \textit{Tool Integration:} Retrieval tools, such as a chunk-level retriever and a document-level retriever, are loaded to support multi-granular semantic search.
\end{itemize}

\paragraph{Document Indexing and Embedding Generation.}
Patent documents are processed and indexed to enable efficient retrieval. Let \( \mathcal{D} \) denote the collection of patent documents and let \( \phi: \mathcal{D} \rightarrow \mathbb{R}^d \) be an embedding function that maps documents to a \( d \)-dimensional vector space. The document embedding process is formalized as: \(\phi(d), \quad \forall d \in \mathcal{D} \). A dense vector representation is computed for each document using a pre-trained embedding model from HuggingFace (e.g., \texttt{intfloat/multilingual-e5-base}), enabling hybrid search that combines semantic similarity and keyword matching.

\paragraph{Reasoning and Retrieval.}
Given a user query \( q \in \mathcal{Q} \), the agent computes its embedding \( \phi(q) \in \mathbb{R}^d \), and employs a ReAct-based chain-of-thought reasoning framework to generate intermediate reasoning steps. Let \( S = \{ s_1, s_2, \dots, s_k \} \) denote the sequence of reasoning steps. 

The agent then chooses a retriever based on the granularity of the query and the reasoning steps generated by the ReAct framework. For instance, if the user is looking for a piece of detailed information, the agent will call the chunk-level retriever, and if the user query requires the integration of information across documents, the agent will call the document-level retriever. The chosen retriever \( R \) will then retrieve a set of documents (or document chunks) from \( \mathcal{D} \) based on a similarity measure such as cosine similarity:
\begin{equation}
R(q) = \{ d \in \mathcal{D} \mid \cos(\phi(q), \phi(d)) \geq \tau \},
\end{equation}
where \( \tau \) is a predefined similarity threshold. The cosine similarity is defined as:
\begin{equation}
\cos(\phi(q), \phi(d)) = \frac{\langle \phi(q), \phi(d) \rangle}{\|\phi(q)\| \, \|\phi(d)\|}.
\end{equation}

 The final response is synthesized by a generation function \( G \) that integrates the query, the reasoning chain, and retrieval results: \( r = G\big(S, R(q), q\big) \). Here, \( G \) encapsulates the output parsing and final response generation, ensuring that the answer is coherent and contextually grounded.

\paragraph{Interactive Query Handling.}
The QA agent is integrated with a Gradio-based user interface that supports real-time interactions. User queries, along with maintained chat history \( H \), are passed to the agent. The overall system behavior can be summarized as a composite function: \( r = \big( G \circ F \circ \phi \big)(q, H) \), where \( \phi \) computes the embedding for \( q \), \( F \) represents the combined chain-of-thought reasoning and retrieval operations, \( G \) generates the final response.

\paragraph{Overall Workflow.}
The end-to-end operation of the QA agent proceeds as follows:
\begin{itemize}[nosep,leftmargin=*]
    \item[$\bullet$]  \textbf{Preprocessing and Indexing:} Patent documents are parsed, embedded, and indexed in the Neo4j vector database.
    \item[$\bullet$] \textbf{ReAct-Based Reasoning and Retrieval:} Upon receiving a user query \( q \), the LLM, guided by a custom prompt and incorporating previous chat history \( H \), generates a chain-of-thought \( S \) that leads to intermediate tool invocations, and then retrieves relevant documents \( R(q) \).
    % \item[$\bullet$] \textbf{Document Retrieval:} Upon receiving a user query \( q \), the system computes \( \phi(q) \) and retrieves relevant documents \( R(q) \).
    \item[$\bullet$] \textbf{Response Synthesis:} The final answer is synthesized as \( r = G(S, R(q), q) \) and returned to the user.
\end{itemize}

This integrated approach, combining structured LLM reasoning with robust embedding-based retrieval, enables efficient and context-aware patent information retrieval.

\section{Experiments: Patent Knowledge Graph Construction: From OCR+LLM to VQA}
We evaluate two methods of constructing patent knowledge graphs by extracting structured patent information from patent documents: an OCR+LLM pipeline and a Visual Question Answering (VQA) based approach. Our experiments focus on the systems' ability to accurately and efficiently extract key information including patent number, patent name, applicant, inventors, assignee, cited patents, and classification fields.

\subsection{Experimental Setup}
The experiments use a dataset of PDF-formatted patent documents obtained from universities' internal patent database\footnote{These patents are all publicly available in the USPTO database, but in order to avoid any leakage of authors' institutions, we will not release this dataset at the current stage.}. This dataset includes patents with varying layouts and levels of complexity.

The OCR+LLM pipeline first applies optical character recognition (OCR) to extract text from the patent cover image. The extracted text is then processed by a large language model, which organizes the information into a structured knowledge graph representing the key patent details.  In contrast, the VQA-based approach directly leverages visual reasoning by combining the cover image with a textual query, making the intermediate OCR text extraction step unnecessary. These two methods are tested only on the cover pages of patent documents, considering that the cover pages already contain most of the key information of our interest. Both methods were deployed under comparable hardware settings with GPU acceleration (CUDA) when available, ensuring a fair comparison of processing efficiency and robustness against document variability.

\subsection{Results and Discussion}
The quality of the constructed KGs were evaluated on carefully selected metrics. First, extraction time per patent was calculated to assess the generation speed. Secondly, accuracy of extracted patent information and ratio of misclassified patents were evaluated, under the condition that all the patents were sampled from the same applicant organization. These metrics can effectively assess the efficiency and reliability of methods of constructing patent KGs.

Preliminary experimental results demonstrate that the VQA-based method significantly outperforms the OCR+LLM pipeline. The VQA approach exhibits enhanced robustness in handling complex cover layouts and minimizes errors associated with OCR noise. In contrast, OCR damages the layout of the original patent document, and the sequential nature of the OCR+LLM pipeline leads to error propagation to the LLM inference stage.

\begin{table}[t]
    \centering
    {\fontsize{9.7pt}{11.0pt}\selectfont 
    \renewcommand{\arraystretch}{1.5}
    \begin{tabular}{l|ccc}
        \hline
        Model & Time (s) & RAE  & RIC \\
        \hline
        Qwen2-7B         & 6.54  & 63.07\% & 12.92\% \\  
        \textbf{Qwen2-VL-7B}    & \textbf{5.05}  & \textbf{82.11\%} & \textbf{9.58\%} \\
        \hline
        Qwen2.5-7B       & 9.80  & 78.50\% &  15.04\% \\
        \textbf{Qwen2.5-VL-7B}  & \textbf{8.38}  & \textbf{92.35\%} & \textbf{7.31\%} \\
        \hline
    \end{tabular}
    }
    \caption{Comparison of OCR+LLM and VQA for knowledge graph construction. Metrics (as detailed in \Cref{sec:metrics}) include average extraction time, RAE (Ratio of Accurately Extracted Entities), and RIC (Ratio of Incorrectly Classified Clusters). All patents originate from the same organization.
    }
    \label{tab:experiment}
\end{table}

\begin{figure}[t]
  \centering  
      \begin{subfigure}[b]{0.49\linewidth}
        \centering
        \includegraphics[width=\linewidth]{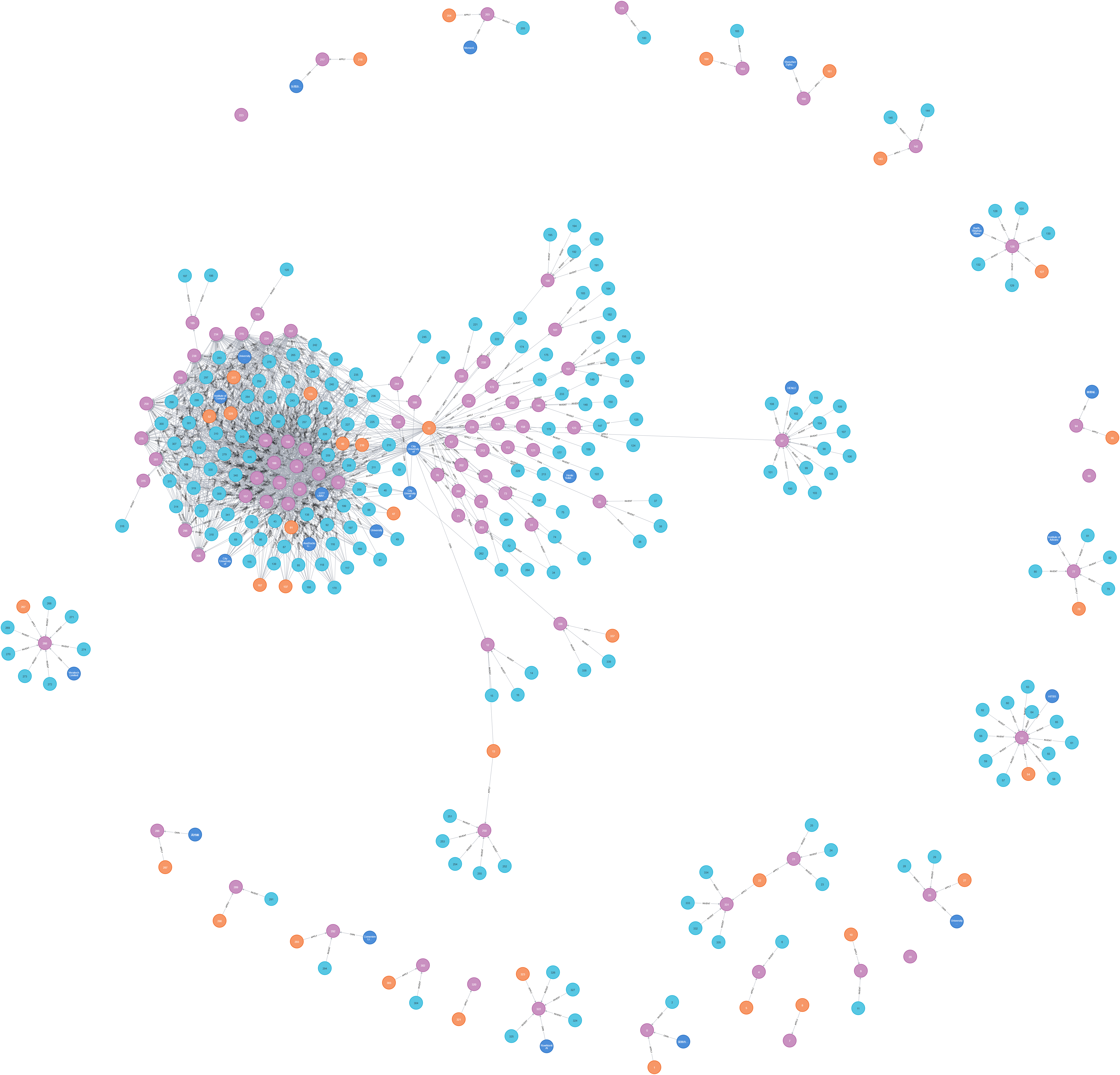}
        \caption{OCR+Qwen2.5-7B}
    \end{subfigure}
    \hfill
    \begin{subfigure}[b]{0.49\linewidth}
        \centering
        \includegraphics[width=\textwidth]{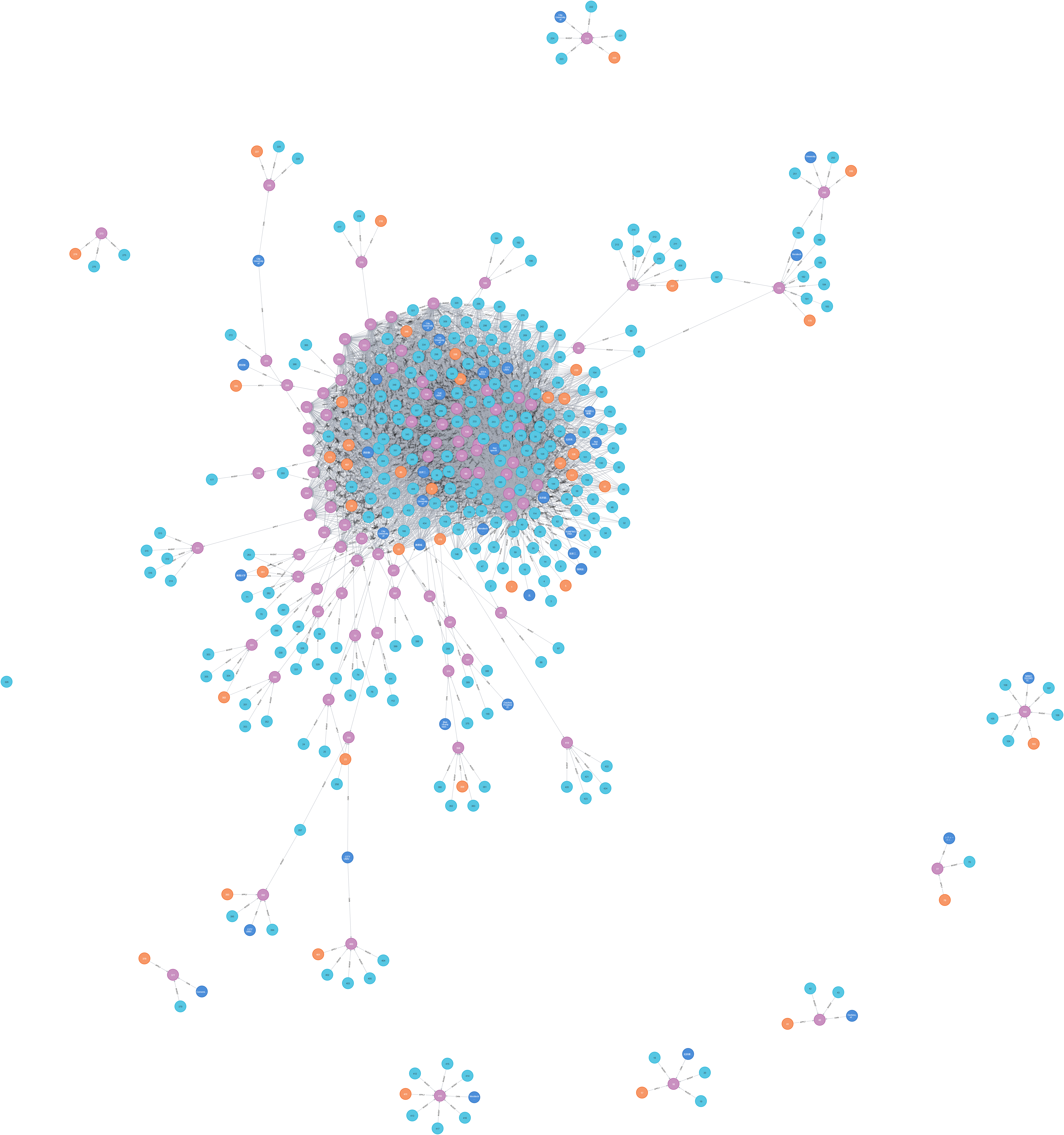}
        \caption{Qwen2.5-VL-7B VQA}
    \end{subfigure}
  \caption{Patent knowledge graphs from OCR+LLM and VQA. The VQA method produces denser entity connections, demonstrating superior information extraction and relationship identification.}
  \label{fig:experiment}
\end{figure}

As shown in \Cref{tab:experiment} and \Cref{fig:experiment}, our findings indicate that directly integrating visual information through VQA not only facilitates extraction speed but also improves extraction accuracy. These results suggest that image-based reasoning is promising for reliable patent information retrieval.

\section{Summary, Conclusion and Future Work}
This work presents KLIPA, a knowledge graph and LLM-driven question-answering framework tailored for patent retrieval and analysis. By integrating structured knowledge graphs, LLM agent-powered reasoning, and retrieval-augmented generation, KLIPA enables more efficient, accurate, and adaptive patent information retrieval. The knowledge graph organizes structured entity relationships, RAG identifies latent semantic connections beyond explicit links, and the LLM agent dynamically optimizes query execution, improving the efficiency of patent search and knowledge discovery.

Evaluations on a university patent dataset show significant improvements in retrieval accuracy, response relevance, and automation efficiency, reducing reliance on domain experts while streamlining intellectual property analysis.

Future work includes expanding multilingual support, introducing advanced reasoning methods like causal inference, and deploying KLIPA in legal and patent office environments to refine the system for practical use.

\section*{Limitations}
Despite the strengths of our proposed system, several limitations remain:
\begin{itemize}[nosep,leftmargin=*]
\item[$\bullet$] Scalability: As the knowledge graph grows, maintenance and updates become more resource-demanding, requiring optimized indexing and storage strategies.
\item[$\bullet$] Data Quality Dependence: The system’s accuracy can be influenced by inconsistencies or missing metadata in patent records.
\item[$\bullet$] Complex Queries: While the hybrid search improves relevance, highly intricate queries may still require domain-specific fine-tuning.
\end{itemize}

Addressing these limitations in future iterations of our system will be crucial for enhancing KLIPA's robustness, accuracy, and applicability across broader intellectual property and legal domains.

\section*{Ethics Statement}
We all comply with the ACM Code of Ethics\footnote{\url{https://www.acm.org/code-of-ethics}} during our study. All datasets used contain anonymized consumer data, ensuring strict privacy protections.

\bibliography{custom}

\appendix

\section{Metrics}
\label{sec:metrics}

This section provides detailed definitions for the metrics used to evaluate the performance of our knowledge graph construction methods.

\subsection{RAE (Ratio of Accurately Extracted Entities)}

The Ratio of Accurately Extracted Entities (RAE) is a metric designed to measure the accuracy of the information extraction process. It is calculated as the proportion of correctly identified and extracted entities relative to the total number of entities that should have been extracted from the patent document.

An \textit{entity} refers to a key piece of structured information on the patent's cover page, including but not limited to:
\begin{itemize}
    \item Patent Number
    \item Patent Name
    \item Applicant(s)
    \item Inventor(s)
    \item Assignee
    \item Cited Patents
    \item Classification Fields
\end{itemize}

The formula is defined by dividing the number of accurately extracted entities ($N_{\text{accurate}}$) by the total number of ground truth entities ($N_{\text{total}}$):
\begin{equation}
    \text{RAE} = \frac{N_{\text{accurate}}}{N_{\text{total}}} \times 100\%
\end{equation}

A higher RAE value indicates a more accurate and reliable extraction performance, signifying that the model is proficient at correctly identifying and capturing individual data points from the source document.

\subsection{RIC (Ratio of Incorrectly Classified Clusters)}

The Ratio of Incorrectly Classified Clusters (RIC) evaluates the model's ability to correctly recognize the underlying relationships between patents. This metric is particularly relevant given the experimental setup, where all patents used for knowledge graph construction were sourced from the same applicant organization.

Ideally, the resulting knowledge graph should represent all these patents as a single, densely connected central cluster, reflecting their common origin. The RIC is defined as the percentage of patents that the model fails to associate with this main organizational cluster.

The formula is calculated by dividing the number of misclassified patents ($P_{\text{misclassified}}$) by the total number of patents ($P_{\text{total}}$):
\begin{equation}
    \text{RIC} = \frac{P_{\text{misclassified}}}{P_{\text{total}}} \times 100\%
\end{equation}

A lower RIC value is desirable, as it signifies a superior capability of the model to identify and correctly represent the shared institutional affiliation among the patents, leading to a more coherent and logically structured knowledge graph.

\section{Implementation Details}
\label{section: implementation}
This appendix presents implementation details of constructing the patent knowledge graph and the LLM-driven QA assistant.

\subsection{Patent KG Construction}
To perform the multiple phases stated in Section~\ref{sec:patent kg}, we implement two methods to extract information from patent documents: an OCR+LLM pipeline and a Visual Question Answering (VQA) based approach. For the OCR+LLM pipeline, large language models (LLMs) are used to process the text obtained via OCR, while for the VQA-based method, visual language models (VLMs) are implemented to read related information directly from the patent documents. Table~\ref{tab:model size} provides details of the models used when implementing these two methods. 

% 在这里写：专利的dataset多大，KG entity个数，平均多少个relation，

\subsection{LLM-Driven QA Assistant}
To perform the multiple phases stated in Section~\ref{sec:patent qa}, we use an open-source LLM to generate intermediate reasoning steps and final outputs, an open-source embedding model to convert text into vector representations, and a vector database to store the vectorized patent documents. Table~\ref{tab:qa agent components} provides details of these key components of the QA agent. 

\subsection{Experiment Environment}
All experiments have been run on the following hardware:
\begin{itemize}[nosep,leftmargin=*]
    \item[$\bullet$] \textbf{OS:} Ubuntu 24.04.1 LTS.
    \item[$\bullet$] \textbf{CPU:} Intel(R) Xeon(R) Gold 6442Y.
    \item[$\bullet$] \textbf{RAM:} 1.0 TiB.
    \item[$\bullet$] \textbf{GPU:} NVIDIA L40S.
    \item[$\bullet$] \textbf{Software:} Python 3.13.0, PyTorch 2.6.0+cu124, CUDA 12.5.
\end{itemize}

\begin{table}[ht]
    \centering
    %\small
    {\fontsize{9.7pt}{11.0pt}\selectfont
    \renewcommand{\arraystretch}{1.5}
    \resizebox{\columnwidth}{!}{%
        \begin{tabular}{l|cc}
            \hline
            Model & Type & Size\\
            \hline
            Qwen2.5-7B-Instruct & Text-to-Text & 7.62B \\  
            Qwen2-7B-Instruct & Text-to-Text & 7.62B \\
            Qwen2.5-VL-7B-Instruct & Image-Text-to-Text & 8.29B \\
            Qwen2-VL-7B-Instruct & Image-Text-to-Text & 8.29B \\
            \hline
        \end{tabular}
    }
    }
    \caption{LLMs used to construct patent KGs. There are two types of LLMs used to construct the patent KGs: one is the text-to-text type (Qwen2.5-7B-Instruct \cite{yang2024qwen2} and Qwen2-7B-Instruct \cite{yang2024qwen2technicalreport}), the other is the image-text-to-text type (Qwen2.5-VL-7B-Instruct \cite{bai2025qwen2} and Qwen2-VL-7B-Instruct \cite{wang2024qwen2vlenhancingvisionlanguagemodels}). The former extracts entity relationships from the text obtained via OCR, while the latter recognizes information directly from the original patent documents.}
    \label{tab:model size}
\end{table}

\begin{table}[ht]
    \centering
    %\small
    {\fontsize{9.7pt}{11.0pt}\selectfont 
    \resizebox{\columnwidth}{!}{%
        \begin{tabular}{l|cc}
            \hline
            Name & Type & Availability \\
            \hline
            Qwen2.5-7B-Instruct & LLM & Open-Source\\  
            Multilingual-e5-Base & Embedding Model & Open-Source\\
            Neo4j & Vector Database & Open-Source\\
            \hline
        \end{tabular}
    }
    }
    \caption{Key components of the QA Assistant. For the LLM, we choose Qwen2.5-7B-Instruct, an open-source model that is widely used in the production environment. For the embedding model, we choose Multilingual-e5-Base. For the vector database, we choose Neo4j, the same as we use for the construction of patent KGs.}
    \label{tab:qa agent components}
\end{table}

\lstset{
    basicstyle=\ttfamily\small,
    numbers=left,
    numberstyle=\tiny,
    stepnumber=1,
    numbersep=10pt,
    backgroundcolor=\color{white},
    showspaces=false,
    showstringspaces=false,
    showtabs=false,
    frame=single,
    tabsize=2,
    captionpos=b,
    breaklines=true,
    breakatwhitespace=false,
    escapeinside={\%*}{*)}
}

% \section{Metrics}

\section{Pseudocode and Implementation Exemplars} 
\label{section: pseudocode}
This appendix presents technical implementation patterns through executable pseudocode and operational examples from our knowledge graph construction pipeline.

\subsection{Patent KG Generation}
\subsubsection{Data Preprocessing Patterns} \label{ssec:preproc}

\paragraph{Context-Aware Text Segmentation.}
Listing~\ref{lst:splitter} shows an example of text segmentation with overlap management, optimized for BERT-style models.

\begin{figure*}[ht]
\begin{lstlisting}[language=Python, caption={Text segmentation with overlap management.}, label=lst:splitter]
class PatentSplitter(RecursiveCharacterTextSplitter):
    def __init__(self):
        super().__init__(
            chunk_size=200,      # Optimal for BERT-style models
            chunk_overlap=30,    # 15% contextual carryover
            separators=[
                r"(?<=\.)\s*",   # Sentence boundaries
                r"\n\s*\n",      # Paragraph breaks
                r"(?<=\})\s*",   # Document structure markers
            ],
            is_separator_regex=True
        )

    def split_document(self, doc: Document) -> List[Document]:
        chunks = super().split_text(doc.page_content)
        return [
            Document(
                page_content=chunk,
                metadata={**doc.metadata, "seq_id": i}
            )
            for i, chunk in enumerate(chunks)
        ]
\end{lstlisting}
\end{figure*}

\paragraph{Operational Example:}  
\noindent

\textbf{Input:}  
\begin{lstlisting}
"Example 1: Disperse carbon nanotubes (CNT) in ethanol via ultrasonic treatment for 40 minutes..."
\end{lstlisting}

\textbf{Output:}  
\begin{lstlisting}
[ "Example 1: Disperse carbon nanotubes (CNT)...ultrasonic treatment",
  "Ultrasonic treatment for 40 minutes, followed by centrifugation..." ]
\end{lstlisting}

\subsubsection{Relation Extraction Workflows} \label{ssec:relext}

\paragraph{Dynamic Prompt Templating.}
Listing~\ref{lst:prompt} demonstrates adaptive prompt generation for extracting structured information from patent texts.

\begin{figure*}[ht]
\begin{lstlisting}[language=Python, caption={Adaptive prompt generation.}, label=lst:prompt]
def build_prompt(doc: Document, config: SchemaConfig) -> str:
    template = """
    Extract from patent text (metadata: {metadata}):
    {text}
    
    Constraints:
    - Entities: {entities}
    - Relations: {relations}
    - Output format: {format}
    
    Respond ONLY with valid JSON."""
    
    return template.format(
        metadata=doc.metadata.get('source', ''),
        text=doc.page_content[:500] + "...",  # Truncate long texts
        entities=config.entity_types,
        relations=config.relation_matrix,
        format=json.dumps(config.output_schema)
    )
\end{lstlisting}
\end{figure*}

\paragraph{Dual-Stage Parsing Implementation.}
Listing~\ref{lst:parsing} provides a robust JSON parsing strategy involving a two-stage approach and schema validation.

\begin{figure*}[ht]
\begin{lstlisting}[language=Python, caption={Robust JSON parsing.}, label=lst:parsing]
def parse_response(response: str) -> List[Triple]:
    # Stage 1: Standard parsing
    try:
        return json.loads(response)
    except JSONDecodeError:
        pass
    
    # Stage 2: Syntax repair
    repaired = json_repair.loads(
        response, 
        skip_json_attributes=True,
        handle_nested_arrays=True
    )
    
    # Validation
    validate_schema(repaired)
    return repaired

def validate_schema(data: List[dict]) -> None:
    required_keys = {"head", "relation", "tail"}
    for item in data:
        if not required_keys.issubset(item.keys()):
            raise InvalidTripleError(f"Missing keys in {item}")
\end{lstlisting}
\end{figure*}

\subsubsection{Graph Construction Strategies} \label{ssec:graphbuild}

\paragraph{Constraint-Driven Node Creation.}
Listing~\ref{lst:neo4j} illustrates node creation in a Neo4j graph database.

\begin{figure*}[ht]
\begin{lstlisting}[language=Python, caption={Neo4j node creation.}, label=lst:neo4j]
def create_material_node(tx, material: Material):
    query = (
        "MERGE (m:Material {cas: $cas}) "
        "ON CREATE SET m += {props} "
        "RETURN id(m)"
    )
    params = {
        "cas": material.cas_number,
        "props": material.properties
    }
    result = tx.run(query, params)
    return result.single()[0]
\end{lstlisting}
\end{figure*}

\paragraph{Batch Processing Optimization.}
Listing~\ref{lst:batch} demonstrates how to optimize graph construction using batched write operations.

\begin{figure*}[ht]
\begin{lstlisting}[language=Python, caption={Batched write operations.}, label=lst:batch]
class Neo4jBatchWriter:
    def __init__(self, batch_size=100):
        self.batch = []
        self.batch_size = batch_size
        
    def add_triple(self, triple: Triple):
        self.batch.append(triple)
        if len(self.batch) >= self.batch_size:
            self.flush()
            
    def flush(self):
        with self.driver.session() as session:
            session.execute_write(self._process_batch, self.batch)
        self.batch = []
        
    def _process_batch(self, tx, batch):
        query = (
            "UNWIND $batch AS item "
            "MERGE (h:Entity {id: item.head}) "
            "MERGE (t:Entity {id: item.tail}) "
            "MERGE (h)-[r:RELATION {type: item.rel}]->(t)"
        )
        tx.run(query, {"batch": batch})
\end{lstlisting}
\end{figure*}

\subsection{Patent QA Systems}
\label{section: pseudocode}
This appendix presents executable pseudocode and implementation patterns for the LLM-based patent QA agent, covering document indexing, retrieval mechanisms, reasoning workflows, and interactive query handling.

\subsubsection{Document Indexing and Embedding} \label{ssec:docindex}

\paragraph{Vector Embedding Generation.}
Listing~\ref{lst:embed} shows an example of generating dense embeddings for patent documents using a transformer-based model.

\begin{figure*}[ht]
\begin{lstlisting}[language=Python, caption={Embedding generation with a transformer model.}, label=lst:embed]
from sentence_transformers import SentenceTransformer

class PatentEmbedder:
    def __init__(self, model_name="intfloat/multilingual-e5-base"):
        self.model = SentenceTransformer(model_name)
        
    def encode(self, text: str) -> List[float]:
        return self.model.encode(text).tolist()
    
    def embed_document(self, doc: Document) -> Dict:
        return {
            "doc_id": doc.metadata["id"],
            "vector": self.encode(doc.page_content),
            "metadata": doc.metadata
        }
\end{lstlisting}
\end{figure*}

\paragraph{Indexing Implementation.}
\noindent

\textbf{Input:}  
\begin{lstlisting}
Document(id="123", page_content="A novel battery electrolyte containing lithium salts...")
\end{lstlisting}

\textbf{Output:}  
\begin{lstlisting}
{ "doc_id": "123", 
  "vector": [0.12, -0.45, 0.78, ...], 
  "metadata": {...} }
\end{lstlisting}

\subsubsection{Retrieval Mechanisms} \label{ssec:retrieval}

\paragraph{Hybrid Search Strategy.}
Listing~\ref{lst:retrieval} demonstrates a hybrid retrieval method combining semantic similarity and keyword search.

\begin{figure*}[ht]
\begin{lstlisting}[language=Python, caption={Hybrid retrieval mechanism.}, label=lst:retrieval]
class HybridRetriever:
    def __init__(self, vector_store, keyword_index):
        self.vector_store = vector_store
        self.keyword_index = keyword_index
        
    def retrieve(self, query: str, top_k=5) -> List[Document]:
        # Compute query embedding
        query_vec = self.vector_store.embedder.encode(query)
        
        # Retrieve using vector similarity
        vector_results = self.vector_store.search(query_vec, top_k)
        
        # Retrieve using keyword match
        keyword_results = self.keyword_index.search(query, top_k)
        
        # Merge results with weighted ranking
        return self.rank_results(vector_results, keyword_results)

    def rank_results(self, vector_results, keyword_results):
        combined = {doc.id: doc for doc in vector_results + keyword_results}
        return sorted(combined.values(), key=lambda d: d.score, reverse=True)
\end{lstlisting}
\end{figure*}

\subsubsection{LLM Reasoning and Response Synthesis} \label{ssec:reasoning}

\paragraph{ReAct-Based Reasoning Framework.}
Listing~\ref{lst:react} illustrates a reasoning pipeline that generates intermediate thoughts before invoking retrieval tools.

\begin{figure*}[ht]
\begin{lstlisting}[language=Python, caption={Chain-of-thought reasoning with tool invocation.}, label=lst:react]
class ReActAgent:
    def __init__(self, llm, retriever):
        self.llm = llm
        self.retriever = retriever
        
    def generate_reasoning_steps(self, query: str) -> List[str]:
        prompt = f"Given the patent-related query: '{query}', reason step-by-step before answering."
        response = self.llm.generate(prompt)
        return response.split("\n")  # Split into reasoning steps
    
    def execute(self, query: str):
        steps = self.generate_reasoning_steps(query)
        retrieved_docs = self.retriever.retrieve(query)
        
        prompt = f"Using the documents {retrieved_docs}, synthesize a response."
        return self.llm.generate(prompt)
\end{lstlisting}
\end{figure*}

\subsubsection{Interactive Query Handling} \label{ssec:interaction}

\paragraph{LLM Integration with Gradio.}
Listing~\ref{lst:gradio} presents an implementation of a QA interface using Gradio.

\begin{figure*}[ht]
\begin{lstlisting}[language=Python, caption={Interactive patent QA system.}, label=lst:gradio]
import gradio as gr

class PatentQAInterface:
    def __init__(self, qa_agent):
        self.qa_agent = qa_agent
        
    def query_response(self, user_query: str) -> str:
        return self.qa_agent.execute(user_query)

    def launch(self):
        iface = gr.Interface(
            fn=self.query_response,
            inputs="text",
            outputs="text",
            title="Patent QA Agent"
        )
        iface.launch()
\end{lstlisting}
\end{figure*}

\section{QA Examples about Patents}

In this section, we present several QA examples. As shown in \Cref{tab:QAexample}, example 1 demonstrates our system's capability in generating patent summarization, while example 2 showcases its ability to generate detailed patent information.

\begin{table*}[ht]
    \centering
    
    \begin{tabular}{@{}p{0.2\textwidth} p{0.8\textwidth}@{}}
        \toprule
        Question & Answer \\ \midrule
        Write a report about apparatus for facilitating a photovoltaic device to provide a wireless communication channel. &  
        
        Solar Panels and Photovoltaic Modules

(1) Introduction:
Solar panels and photovoltaic (PV) modules are essential for capturing solar energy, designed for high energy conversion efficiency. These modules are often integrated into wireless communication devices and can be mounted on rooftops, towers, or other suitable locations to ensure optimal sunlight exposure.

(2) Energy Harvesting Circuits:
Energy harvesting circuits convert the electrical energy from PV cells into usable power for wireless devices. These circuits include charge controllers, batteries, and voltage regulators. Charge controllers manage the charging process, while voltage regulators ensure stable output voltage for proper device operation.

(3) Power Management Systems:
Power management systems optimize energy usage in wireless communication devices. They include microcontrollers, power switches, and load management units. Microcontrollers monitor energy supply and demand, adjusting device operation accordingly. Power switches manage the energy flow to different components to ensure efficient utilization.

(4) Antenna and Modulation Techniques:
Antennas play a key role in transmitting and receiving signals, with design and placement optimized for maximum signal strength and coverage. Modulation techniques such as frequency modulation (FM) and phase-shift keying (PSK) encode data onto carrier signals, ensuring reliable data transmission.

(5) Applications:
The integration of PV devices with wireless communication systems enables several applications across various industries: Remote Sensing and Monitoring: PV-powered wireless sensors for environmental monitoring, agriculture, and infrastructure health;
Smart Grids: PV-powered communication nodes that enhance smart grid systems by providing real-time data on energy consumption;
Military and Defense: Self-sustaining communication systems for remote military operations and disaster response;
IoT and M2M Communications: PV-powered devices enabling IoT and M2M networks in challenging environments.

(6) Conclusion:
Integrating photovoltaic devices with wireless communication systems offers a sustainable and energy-efficient solution. These systems reduce dependence on non-renewable resources and minimize environmental impact. Continued technological advancements promise further improvements and innovations in this field.

(7) References:
Martin Garaj and Shu Hung Henry Chung. Apparatus for facilitating a photovoltaic device to provide a wireless communication channel. US Patent 11,671,731 B2, 2023.

This report provides an overview of the key components and applications of photovoltaic-integrated wireless communication systems, emphasizing their role in creating efficient, sustainable communication networks.

        \\ \midrule
        What is the field of the invention titled "High-efficiency quantitative control of mitochondrial transfer based on droplet microfluidics"? & The field of the invention titled "High-efficiency quantitative control of mitochondrial transfer based on droplet microfluidics" is biotechnology, particularly in the areas of cell biology and microfluidics. \\ \bottomrule
    \end{tabular}
    \caption{QA Examples.}
    \label{tab:QAexample}
\end{table*}

\end{document}